\RequirePackage{fix-cm}

\documentclass[referee]{svjour3}

\smartqed

\usepackage{amssymb}
\usepackage{amsmath}
\usepackage{graphicx}
\usepackage{caption}

\journalname{Nonlinear Dynamics}

\begin{document}

\title{Duffing revisited: Phase-shift control and internal
resonance in self-sustained oscillators}
\titlerunning{Duffing revisited: Self-sustained oscillators}

\author{Sebasti\'an I. Arroyo \and Dami\'an H. Zanette}

\institute{S. I. Arroyo \at
              Consejo Nacional de Investigaciones
Cient\'{\i}ficas y T\'ecnicas, Universidad Nacional de Quilmes,
Roque S\'aenz Pe\~na 352, 1876 Bernal,
Buenos Aires, Argentina\\
              \email{seba.arroyo7@gmail.com}
           \and
           D. H. Zanette \at
              Consejo Nacional de Investigaciones
Cient\'{\i}ficas y T\'ecnicas, Centro At\'omico Bariloche and
Instituto Balseiro, 8400 San Carlos de Bariloche, R\'{\i}o Negro, Argentina \\
              Tel.: +54-294-4445173\\
              Fax: +54-294-4445299\\
              \email{zanette@cab.cnea.gov.ar}
}

\date{Received: date / Accepted: date}

\maketitle

\begin{abstract}
We address two aspects of the dynamics of the forced Duffing
oscillator which are relevant to the technology of micromechanical
devices and, at the same time, have intrinsic significance to the
field of nonlinear oscillating systems. First, we study the
stability of periodic motion when the phase shift between the
external force and the oscillation is controlled --contrary to the
standard case, where the control parameter is the frequency of the
force. Phase-shift control is the operational configuration under
which self-sustained oscillators --and, in particular,
micromechanical oscillators-- provide a frequency reference useful
for time keeping. We show that, contrary to the standard forced
Duffing oscillator, under phase-shift control oscillations are
stable over the whole resonance curve.  Second, we analyze a model
for the internal resonance between the main Duffing oscillation mode
and a higher-harmonic mode of a vibrating solid bar clamped at its
two ends. We focus on the stabilization of the oscillation frequency
when the resonance takes place, and present preliminary experimental
results that illustrate the phenomenon. This synchronization process
has been proposed to counteract the undesirable frequency-amplitude
interdependence in nonlinear time-keeping micromechanical devices.
\end{abstract}

\section{Introduction} \label{intro}

Since the 1970s, the Duffing oscillator --first introduced by the
German engineer Georg Duffing in 1918 \cite{GDuffing}-- has been
repeatedly invoked as a paradigm for nonlinear behavior in classical
mechanical systems \cite{Duff}. The mere addition of a cubic force
to a linear damped oscillator deploys a plethora of complex
phenomena, but still preserves the simple mathematical form and low
dimensionality that makes the system readily amenable to
computational exploration and, to a certain extent, to analytical
study. Under the action of an external harmonic force, the Duffing
oscillator's response can be bistable, with two possible oscillation
amplitudes for each frequency. This property, which is directly
related to the well-known Duffing leaning resonance curve
\cite{Landau}, has been exploited as an illustration of catastrophe
theory \cite{Holmes1}. If, moreover, the sign of the linear force is
inverted, creating a double-well potential,  the dynamics can turn
into non-periodic motion, with a strange attractor that became a
traditional example of low-dimensional chaos \cite{Holmes2,Holmes3}.

Much more recently, the Duffing oscillator acquired relevance in the
realm of microtechnologies \cite{mm1,mm2,mm3}. It has since long
been known that the Duffing equation stands for the leading
nonlinear correction to the oscillations of an elastic beam clamped
at its two ends \cite{beam1,beam2}. Minute vibrating silica beams,
in turn, have been proposed as pacemakers for the design of
time-keeping devices at the microscale, where traditional quartz
crystals are difficult to build and operate \cite{mm1}. Since, to
overcome the effect of noise, these microscopic silica beams must
vibrate at relatively large amplitudes, their oscillations take
place within the nonlinear regime and --in clamped-clamped
configurations \cite{mm2}-- the appropriate mathematical description
is thus given by the Duffing equation.

\begin{figure} \includegraphics[width=0.4\textwidth]{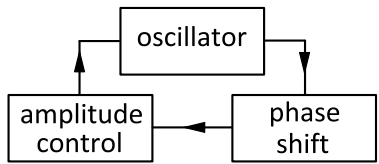}
\caption{Schematic representation of the feedback process that stirs
self-sustained oscillations. The electric signal produced by the
oscillatory motion is conditioned by shifting its phase and
adjusting its amplitude, and is then reinjected as a mechanical
force acting on the oscillator itself.} \label{fig1}
\end{figure}

In order to function as a pacemaker, a mechanical system must be
able to sustain stationary periodic motion with an autonomously
generated frequency. This can be  achieved in practice by inserting
the oscillator into a feedback electronic circuit \cite{feedback}
which, first, reads the signal generated by the oscillator's
displacement from equilibrium. This signal is then conditioned by
shifting its phase by a prescribed amount (or, equivalently, by
inserting a time delay) and fixing its amplitude. Once conditioned,
the signal is transformed into a mechanical force and reinjected to
act on the oscillator. The oscillator, in turn, responds to this
action as an ordinary resonator, except that the ``external'' force
possesses the frequency generated by the oscillator itself.  The
result of this feedback process, schematized in Fig.~\ref{fig1}, is
that the mechanical system reaches self-sustained periodic motion,
with the only external input of the power needed to condition the
electric signal. The frequency of the oscillations is determined by
the mechanical properties of the oscillator and the parameters of
signal conditioning. From the dynamical viewpoint, this
self-sustained configuration has the interest that the control
parameter, to which the experimenter has access at the conditioning
stage,  is the phase shift between the oscillation and the force
--instead of the frequency, as in the standard case of an externally
forced oscillator.

In this paper, we focus on two aspects of the dynamics of the
self-sustained Duffing oscillator which, apart from their interest
from a theoretical perspective,  have specific implications in the
technological applications of the  system. In both cases, we provide
approximate analytical results and a numerical validation. After
briefly reviewing the mathematical model and its main properties in
Section \ref{D}, we first analyze the stability of oscillatory
motion under controlling the phase shift between the oscillations
and the self-sustaining force (Section \ref{FS}). In contrast with
the case where the frequency is controlled, where both stable and
unstable solutions are found for a given set of parameters, we find
that in our case oscillations are always stable. Then, in Section
\ref{IR}, we analyze a model for the coupling between the main
oscillation mode and a higher-harmonic linear mode in a
clamped-clamped oscillator.  The ensuing internal resonance, with
mutual synchronization of the two modes, has been invoked as a
possible method to neutralize the undesirable dependence of the
frequency on the amplitude, characteristic of any nonlinear
oscillating system \cite{mm2}. We present an experimental
demonstration of the internal resonance under phase-shift control,
and fit the experimental results with a simple analytical
approximation to the model. Our conclusions are drawn in Section
\ref{C}.

\section{The self-sustained Duffing oscillator} \label{D}

In its main oscillation mode, an elastic beam  clamped at its two
ends vibrates transversally, much like a plucked string  \cite{mm2}.
For moderate amplitudes,  the displacement from equilibrium  is well
described by a coordinate $x(t)$ satisfying
\begin{equation} \label{adim}
\ddot x + Q^{-1} \dot x + x + \beta x^3 = f_0 \cos (\phi +  \phi_0).
\end{equation}
This equation of motion has been normalized by the effective mass,
and time units have been chosen in such a way that the frequency of
undamped ($Q^{-1}=0$), harmonic ($\beta=0$), unforced ($f_0=0$)
oscillations is equal to one. The quality factor $Q$ gives the ratio
between the typical damping time and the oscillation period --or,
equivalently, between the width of the resonance curve and the
oscillation frequency-- and $\beta$ weights the relative strength of
the nonlinear forcing. Clamped-clamped oscillators have $\beta
>0$, so that the nonlinearity hardens the total force.

The right-hand side of Eq.~(\ref{adim}) stands for the
self-sustaining force provided by the feedback circuit, as described
in the Introduction. Here, $\phi (t)$ is the phase associated to the
oscillatory motion, while $\phi_0$ and $f_0$ are the phase shift and
the amplitude fixed by signal conditioning (see Section
\ref{intro}). For harmonic oscillations, the phase is defined in
terms of the coordinate, from the identity $x(t)=A \cos \phi(t)$.
For other kinds of motion, the phase can be defined in a variety of
ways \cite{kura}. Elsewhere, we have discussed analytical and
numerical definitions which are especially adapted to our specific
problem \cite{PRE}. In the present contribution, however, we deal
mainly with harmonic motion.

\begin{figure} \includegraphics[width=0.75\textwidth]{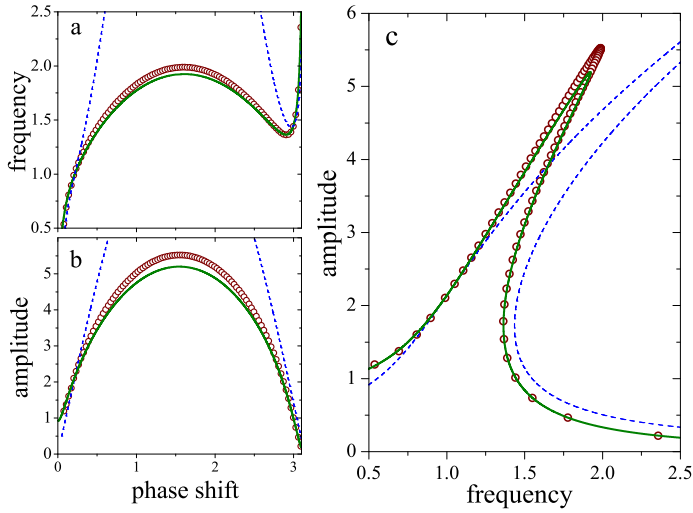}
\caption{Interdependence of amplitude, frequency, and phase shift
for periodic motion of a forced Duffing oscillator, with
$Q^{-1}=0.1$, $\tilde \beta=0.1$, and $f_0=1$. Solid lines:
solutions to Eqs.~(\ref{algeb}); dotted lines: solutions to
Eqs.~(\ref{sfinss}); open dots: long-time numerical solutions to the
equation of motion of the self-sustained oscillator,
Eq.~(\ref{adim}), for numerous values of the phase shift $\phi_0$. }
\label{fig2}
\end{figure}

Assuming that the self-sustained system performs oscillations with
constant amplitude and frequency, the simplest approximation to
handle the nonlinear effects of the cubic term is to disregard
higher-harmonic contributions to the oscillatory motion --as
traditionally done for the forced Duffing oscillator \cite{Landau}.
This corresponds to the lower-order approximation in the harmonic
balance procedure \cite{HB}. To ease the comparison with the
standard forced case, where the right-hand side of Eq.~(\ref{adim})
is replaced by an oscillatory function of prescribed frequency, we
write $\phi+\phi_0 \equiv \Omega t$, so that all phases are measured
from the initial ($t=0$) phase of the self-sustaining force. We
propose a harmonic solution $x(t) = A \cos \phi \equiv A \cos
(\Omega t - \phi_0)$ and neglect higher-harmonic terms by
approximating $\cos^3 \phi = \frac{3}{4} \cos \phi + \frac{1}{4}
\cos 3\phi \approx \frac{3}{4} \cos \phi$. Separating terms
proportional to $\cos \phi$ and $\sin \phi$ yields the algebraic
equations
\begin{equation} \label{algeb}
(1-\Omega^2) A+\tilde \beta A^3 = f_0 \cos \phi_0, \ \ \
Q^{-1}\Omega A = f_0 \sin \phi_0,
\end{equation}
with $\tilde \beta = \frac{3}{4} \beta$. These equations can be
brought to a third-degree polynomial relations between the unknowns,
and are therefore explicitly solvable. Figures \ref{fig2}a and b
show, as solid lines, the solutions for the amplitude $A$ and the
frequency $\Omega$ as functions of the phase shift $\phi_0$, with
$Q^{-1}=0.1$, $\tilde \beta =0.1$, and $f_0=1$. In Fig.~\ref{fig2}c,
the graph of amplitude versus frequency yields the typical Duffing
resonance curve. Over this curve, $\phi_0 \approx 0$ and $\pi$  at
the leftmost and  rightmost ends, respectively. At the peak, where
the oscillator's response to the self-sustaining force is maximal,
we have $\phi_0 = \pi/2$. For this phase shift, in fact, the force
and the velocity $\dot x$ oscillate in-phase.

It is important to realize that the functional interdependence
between $A$, $\Omega$, and $\phi_0$, given by Eqs.~(\ref{algeb}), is
exactly the same whether the oscillator is subject to the
self-sustaining force that fixes the phase shift $\phi_0$, or
whether it moves under the action of an external force of given
frequency $\Omega$. In particular, the resonance curve in
Fig.~\ref{fig2}c is the same irrespectively of the control parameter
being the phase shift or the frequency. In the latter case, it is
well known that, within the frequency range where three solutions
for the amplitude exist, the two outermost solutions are stable,
while the inner solution is unstable \cite{Landau,Nayfeh}. On the
other hand, the frequency and amplitude plots in  Fig.~\ref{fig2}a
and b suggest that, as the phase shift is varied, no bifurcations
take place, so that stability should not change. In the next
section, we show that --in contrast with the standard forced Duffing
oscillator-- periodic motion under the action of a self-sustaining
force with phase-shift control is in fact stable over the {\em
whole} resonance curve.

\section{Stability under phase-shift control} \label{FS}

Whether the steady oscillations described by Eqs.~(\ref{algeb}) are
stable, and represent the asymptotic motion of the Duffing
oscillator, can be decided by a variety of perturbation techniques
which, as a byproduct, provide an approximate solution to
Eq.~(\ref{adim}) beyond stationary harmonic oscillations. In our
case, a convenient approach is provided by the method of multiple
time scales \cite{Nayfeh}, which --as demonstrated below-- allows
the oscillation frequency to vary in response to the control of the
phase shift, and vice versa. This method discerns between a scale
typical of the oscillatory motion, with period of order unity, and a
longer time scale characteristic of slow changes in amplitude and
frequency. It thus requires that all forces acting on the
oscillator, apart from the linear elastic interaction, are treated
as perturbations. To such end, we rewrite Eq.~(\ref{adim}) as
\begin{equation} \label{adimpert}
\ddot x + x = -\epsilon \left[ Q^{-1} \dot x+\beta x^3  -f_0 \cos
(\phi + \phi_0) \right],
\end{equation}
with $\epsilon $ the perturbation parameter. We write again
$\phi+\phi_0 \equiv \Omega t$ and, moreover, take $\Omega= 1 +
\epsilon \omega$. Note that, while for the standard forced Duffing
oscillator the phase difference between oscillation and forcing
varies with time as the system approaches its asymptotic motion, in
the self-sustained oscillator this phase shift is fixed and it is
the oscillation frequency (i.e., $\omega$) which changes as time
goes on.

The method of multiple scales assumes that the solution to
Eq.~(\ref{adimpert}), $x(t;\epsilon)$, depends on time through two
auxiliary variables, $\tau_0 \equiv t$ and $\tau_1 \equiv \epsilon
t$, respectively representing the fast and slow scales, and that it
can be expanded as $x (t;\epsilon) = x_0(\tau_0,\tau_1) + \epsilon
x_1 (\tau_0,\tau_1) + \cdots$. Inserting this ansatz, and taking
into account that $\frac{d}{dt}=\frac{\partial}{\partial \tau_0} +
\epsilon \frac{\partial}{\partial \tau_1}$, zeroth and first-order
terms in $\epsilon$ yield
\begin{equation} \label{zeroth}
\frac{\partial^2 x_0}{ \partial \tau_0^2} + x_0 =0,
\end{equation}
and
\begin{equation} \label{first}
\frac{\partial^2 x_1}{ \partial \tau_0^2}+ x_1 =-2 \frac{\partial^2
x_0}{ \partial \tau_0 \partial\tau_1} -Q^{-1} \frac{\partial x_0}{
\partial \tau_0}-\beta x_0^3 + f_0 \cos (\tau_0 + \omega
\tau_1),
\end{equation}
respectively. Here, we have used the fact that $\Omega t =\tau_0 +
\omega \tau_1$. The (real) solution to Eq.~(\ref{zeroth}) is
\begin{equation} \label{zsol}
x_0 (\tau_0,\tau_1) = \frac{1}{2}  a(\tau_1) \exp (i\tau_0) +
\mbox{c.c.} = \operatorname{Re}[  a(\tau_1) \exp (i\tau_0)],
\end{equation}
i.e.~an oscillation modulated by a slowly varying amplitude. This
solution must now be replaced into Eq.~(\ref{first}) to find $x_1
(\tau_0,\tau_1)$. The solution for $x_1$, however, contains terms
which grow indefinitely as time elapses. These secular contributions
disappear if the following condition is required to hold:
\begin{equation}
\frac{1}{2} \tilde \beta a^2 a^* + i \left( a' + \frac{1}{2} Q^{-1}
a\right) = \frac{1}{2} f_0 \exp(i\omega \tau_1),
\end{equation}
where the prime denotes differentiation with respect to $\tau_1$,
and $a^*$ is the complex conjugate  of $a$. This condition amounts
to a differential equation to be satisfied by the complex amplitude
$a(\tau_1)$. Writing $a= A \exp( i\alpha)$, where $A$ and $\alpha$
are real functions of $\tau_1$, we get
\begin{equation} \label{sfin}
\begin{array}{rl}
2A' &= - Q^{-1} A + f_0 \sin (\omega \tau_1-\alpha) ,\\
2A\alpha' &= \tilde  \beta A^3 - f_0 \cos (\omega \tau_1-\alpha).
\end{array}
\end{equation}

Note that, having written $a= A \exp( i\alpha)$,  the zeroth-order
solution for $x(t)$ --given by Eq. (\ref{zsol})-- can be cast as
$x_0 = A \cos [\Omega t - (\omega \tau_1 -\alpha ) ]$. Thus, within
this approximation, the combination $\psi = \omega \tau_1-\alpha$ is
nothing but the phase shift between the coordinate and the forcing.
For the standard forced Duffing oscillator, $\psi$ varies with time
along with the oscillation amplitude $a$, while the frequency of the
external force, $\Omega=1+\epsilon \omega$, remains constant. In
this case, Eqs. (\ref{sfin}) govern the evolution of $a$ and $\psi$
through the relation $\psi' = \omega-\alpha'$. Linearization of the
equations yields the stability properties of their equilibrium
solutions. In particular, this procedure shows the possible
existence of the bistability frequency range commented on at the end
of Section \ref{D} (see Fig.~\ref{fig2}c) \cite{Nayfeh}.

For the self-sustained oscillator, on the other hand, $\psi$
coincides with the fixed phase shift $\phi_0$ determined by the
feedback circuit. Therefore, $\psi' = \omega' \tau_1 + \omega -
\alpha'=0$, and  Eqs. (\ref{sfin}) transform into equations for the
amplitude $a$ and the frequency $\omega$:
\begin{equation} \label{sfinss}
\begin{array}{rl}
2A' &= - Q^{-1} A + f_0 \sin \phi_0 ,\\
2A \omega' &=  \tau_1^{-1} ( \tilde  \beta A^3 - f_0 \cos
\phi_0-2A\omega).
\end{array}
\end{equation}
The stationary solutions to these equations coincide with those of
Eqs.~(\ref{algeb}) up to a term of order $\epsilon$ in the
frequency. They are plotted in Fig.~\ref{fig2} as dotted lines. The
substantial difference with the solutions to Eqs.~(\ref{algeb}),
plotted as solid lines, is to be ascribed to the fact that, for the
chosen values of $Q^{-1}$, $\tilde \beta$, and $f_0$, the effects of
damping, nonlinearities, and external forcing are actually sizable
additions to the linear elastic force, instead of small
perturbations.

\begin{figure} \includegraphics[width=0.6\textwidth]{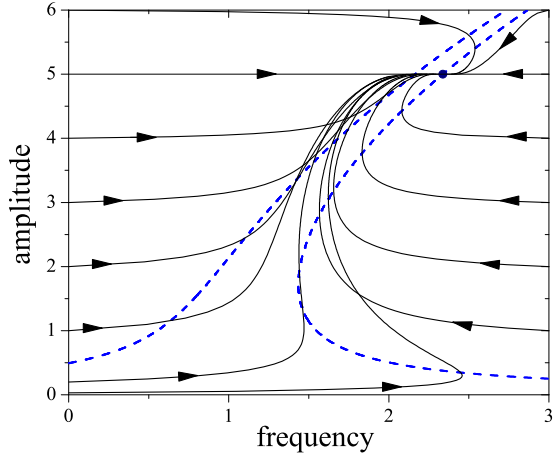}
\caption{The solutions to Eqs.~(\ref{sfinss}) plotted in the
$(\Omega, A)$-plane, for $Q^{-1}=0.1$, $\tilde \beta=0.1$, $f_0=1$,
and $\tau_i=1$. The phase shift for the self-sustaining force  is
fixed to $\phi_0=\frac{5}{6} \pi$. Different lines correspond to
different initial conditions. The full dot stands for the position
of the equilibrium state, $(\frac{9}{4}+\frac{1}{20} \sqrt{3},5)
\approx (2.337,5)$, which lies on the intermediate-amplitude branch
of the resonance curve (dashed line).} \label{fig3}
\end{figure}

To avoid the singularity at $\tau_1=0$ in the second of
Eqs.~(\ref{sfinss}), we integrate the equations from $\tau_1 =
\tau_i >0$. The solutions are
\begin{equation} \label{asol}
A(\tau_1) = Qf_0 \sin \phi_0 +\left[ A(\tau_i)-Qf_0 \sin \phi_0
\right] \exp [-(\tau_1-\tau_i) / 2Q],
\end{equation}
and
\begin{equation} \label{wsol}
\omega (\tau_1) =  \tau_1^{-1} \left[ \tau_i \omega (\tau_i) +
\int_{\tau_i}^{\tau_1} W(\tau) \, d\tau \right],
\end{equation}
with $W (\tau) = \frac{1}{2} A^{-1} (\tau ) [\tilde \beta A^3(\tau)
- f_0  \cos \phi_0]$. Lines with arrows in Fig.~\ref{fig3} show
these solutions in the frequency-amplitude plane, for various
initial conditions, with $Q^{-1}=0.1$, $\tilde \beta=0.1$, $f_0=1$,
and $\tau_i=1$. Note that, because of the dependence of the
right-hand side of the second of Eqs.~(\ref{sfinss}) on $\tau_1$,
the trajectories can cross each other. The phase shift, $\phi_0 =
\frac{5}{6} \pi$, has been chosen in such a way that the
corresponding equilibrium state lies on the branch of the resonance
curve where the standard forced Duffing oscillator is unstable. For
the self-sustained oscillator, in contrast, this equilibrium is
clearly {\sl stable}.

Indeed, Eqs.~(\ref{asol}) and (\ref{wsol}) show that the equilibria
of the self-sustained Duffing oscillator are stable for any value of
the phase shift $\phi_0$, i.e. along the whole resonance curve.
Moreover, for each value of $\phi_0$, the corresponding equilibrium
attracts the trajectories starting at any initial condition and,
hence, it is globally stable. The final approach to the equilibrium
state is exponential in time for the amplitude, while it behaves as
$\tau_1^{-1}$ for the frequency. The much slower long-time
dependence for the frequency explains the flat asymptotic profile of
the trajectories as seen in Fig.~\ref{fig3}.

The validity of the stability analysis based on Eqs.~(\ref{sfinss})
is in principle limited to the perturbative limit, $\epsilon \to 0$.
To decide whether the self-sustained Duffing system exhibits stable
oscillations for any value of the phase shift beyond that limit, we
implemented a numerical integration of the equation of motion
(\ref{adim}). While the equation can be solved by standard
techniques --in our case, the second-order Runge-Kutta algorithm--
the numerical definition of the phase $\phi$ in the self-sustaining
force requires extending the notion of phase to non-oscillatory
behavior. In fact, during the numerical computation of $x(t)$ it
cannot be assumed that the motion is always a harmonic oscillation.
In a previous publication \cite{PRE}, we have given details on a
suitable method to assign an instantaneous phase $\phi(t)$ to any
form of $x(t)$, by locally fitting a trigonometric function. Dots in
Fig.~\ref{fig2} show long-time numerical measurements of amplitude
and frequency for $Q^{-1}=0.1$, $\tilde \beta=0.1$,  $f_0=1$, and
numerous values of $\phi_0$, ranging from $\phi_0\approx 0$ to $\pi$
at intervals of $0.01\, \pi$. As in the perturbative limit, the
solutions over the resonance curve are stable for all $\phi_0$. The
difference with the perturbative stationary solution (dashed line)
shows that we are well beyond that limit. On the other hand, the
solution to Eqs.~(\ref{algeb}) (solid line) --obtained within the
traditional approximation of nonlinearities-- gives a very
satisfactory description of the numerical results, except for the
immediate vicinity of the peak.

\section{Internal resonance in a clamped-clamped oscillator} \label{IR}

As any other elastic  body, the clamped-clamped beam has essentially
an infinite number of oscillation modes. Nonlinear effects can
couple these modes with each other, establishing an interaction in
the form of mutual resonant excitation between different forms of
oscillatory motion. Specifically, cubic nonlinearities make it
possible that oscillations in the main mode synchronize with
higher-harmonic modes whose frequency is three (or a multiple  of
three) times the fundamental frequency.

In a recent experiment \cite{mm2}, a silica clamped-clamped
self-sustained  microoscillator, about $500$ $\mu$m long, $3$ $\mu$m
width, and $10$ $\mu$m thick, was shown to perform oscillations
which, at very small amplitudes, had a frequency of approximately
$66$ kHz. As the amplitude of the self-sustaining force was
gradually increased by the experimenter, the amplitude of the
oscillations  increased as well, and --due to the hardening
nonlinearity of the system-- so did their frequency. When the
frequency attained $68$ kHz, however, it was observed that both the
amplitude and the frequency ceased to grow, and both quantities
reached a wide plateau where they remained practically constant. The
self-sustaining force had to reach more than twice its amplitude at
the beginning of the plateau for the oscillation amplitude and
frequency to regain their growth.

The stabilization of the oscillation amplitude and frequency in the
experiment was attributed to the resonant coupling between the main
oscillation mode, sustained by the feedback circuit, and a
higher-harmonic mode. In fact, finite-element numerical simulations
of the clamped-clamped beam showed the existence of a torsional
oscillation mode with a natural frequency slightly larger than three
times the natural frequency of the main mode. As advanced in the
Introduction, the technological interest of this phenomenon resides
in the fact that an oscillator functioning in the resonant regime
would maintain a very stable frequency, insensible to amplitude
fluctuations in the self-sustaining force, thus providing a more
reliable frequency reference \cite{af}. From a more fundamental
perspective, an interesting aspect of this internal resonance is
that nonlinearities play a twofold role in its origin. First, they
induce the oscillation frequency to vary with the amplitude and, as
a consequence, to reach the value where resonance is possible.
Second, they are the source of the coupling between the main
oscillation mode and higher harmonics.

In the experiment, the internal resonance was brought about by
changing the amplitude $f_0$ of the self-sustained force. The phase
shift in the feedback circuit, on the other hand, was kept constant
at $\phi_0 \approx \pi /2$, i.e.~close to the peak of the resonance
curve, where the oscillator response to self-sustaining was maximal.
Here, we show that the internal resonance can also be induced by
varying the phase shift with fixed $f_0$. To this end, we implement
a variation of the theoretical model used to explain the above
described experimental results \cite{mm2}.

\subsection{Theoretical model} \label{IRth}

In our description, following Sections \ref{D} and \ref{FS}, the
main oscillation mode is represented by a coordinate $x_1(t)$
satisfying the self-sustained Duffing equation (\ref{adim}). For the
sake of simplicity, the higher-harmonic mode is represented by a
coordinate $x_2(t)$ satisfying a {\em linear} unforced oscillator
equation. The two equations are coupled to each other through linear
interactions. Thus, for the main mode we have
\begin{equation} \label{adim1}
\ddot x_1 + Q_1^{-1} \dot x_1 + x_1 + \beta x_1^3 = f_0 \cos (\phi +
\phi_0) + J_1 x_2,
\end{equation}
while for the higher harmonic we have
\begin{equation} \label{adim2}
\omega_2^{-2}  \ddot x_2 +\omega_2^{-1}   Q_2^{-1} \dot x_2 + x_2  =
J_2 x_1,
\end{equation}
where $J_{1,2}$ are the coupling intensities, and $\omega_2$ is the
higher-harmonic frequency (measured in units of the fundamental
frequency; see Eq.~(\ref{adim})).

In order to focus on the phenomenon of frequency stabilization, we
disregard the tripling of the frequency induced by the cubic
nonlinearity, and concentrate on the synchronization of the two
modes, assuming that their frequencies are similar. In other words,
we take $\omega_2 \gtrsim 1$. We thus look for solutions to
Eqs.~(\ref{adim1}) and (\ref{adim2}) where the two modes oscillate
with the same frequency and their phases are locked to each other:
$x_1(t)=A_1 \cos (\Omega t-\phi_0)$, $x_2(t)=A_2 \cos (\Omega
t-\phi_2)$. Proceeding as with Eq.~(\ref{adim}), we find the
following algebraic equations for the amplitudes $A_{1,2}$, the
frequency $\Omega$ and the phase difference $\psi_2=\phi_0-\phi_2$
between the higher-harmonic oscillation and the main mode:
\begin{equation} \label{algebres}
\begin{array}{l}
(1-\Omega^2) A_1 +\tilde \beta A_1^3 = f_0 \cos \phi_0 +J_1A_2 \cos
\psi_2 ,\\
Q_1^{-1}\Omega  A_1 =f_0 \sin \phi_0 +J_1A_2 \sin \psi_2,\\
(\omega_2^2 -\Omega^2) A_2 \cos \psi_2 -\omega_2 Q_2^{-1}\Omega A_2
\sin
\psi_2= \omega_2^2 J_2 A_1 , \\
(\omega_2^2 -\Omega^2) A_2 \sin \psi_2 +\omega_2 Q_2^{-1}\Omega A_2
\cos \psi_2= 0.
\end{array}
\end{equation}
As in the case of Eqs.~(\ref{algeb}), these equations are equivalent
to third-degree polynomial relations between the unknowns, and thus
can be exactly solved.

\begin{figure} \includegraphics[width=0.75\textwidth]{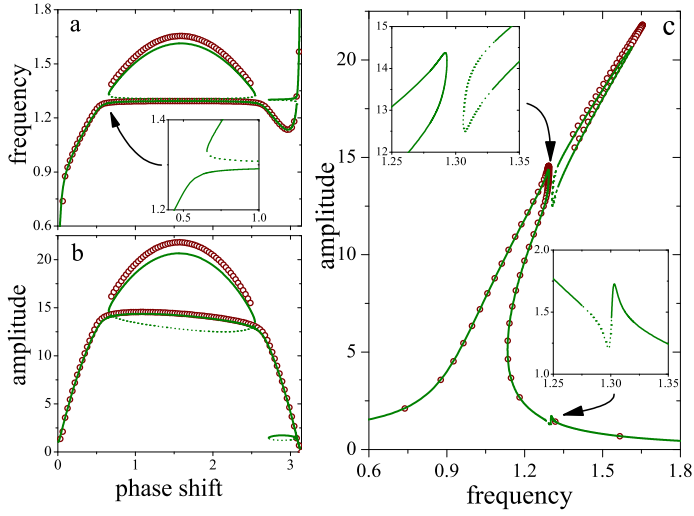}
\caption{Interdependence of the amplitude $A_1$ of the main
oscillation mode, the synchronization frequency $\Omega$, and the
self-sustaining phase shift $\phi_0$, for a forced Duffing
oscillator coupled to a linear oscillator representing a
higher-harmonic mode. Lines: solutions to Eqs.~(\ref{algebres}),
with $ \beta =0.005$, $f_0=1$, $\omega_2=1.3$, $Q_1^{-1}=0.03$,
$\omega_2^{-1} Q_2^{-1} =0.003$, $J_1=10^{-4}$, and $J_2=1$. Solid
and dotted sections represent stable and unstable oscillations,
respectively. Open dots: long-time numerical solutions to
Eqs.~(\ref{adim1}) and (\ref{adim2}) for the same set of parameters,
and numerous values of the phase shift $\phi_0$. The insets show
close-ups of the zones pointed to by the arrows. For clarity,
numerical results are not shown in the insets.} \label{fig4}
\end{figure}

Curves in the three panels of Fig.~\ref{fig4} illustrate the
interdependence between the amplitude $A_1$, the synchronization
frequency $\Omega$, and the phase shift $\phi_0$, as determined by
Eqs.~(\ref{algebres}), for  $ \beta =0.005$, $f_0=1$,
$\omega_2=1.3$, $Q_1^{-1}=0.03$, $\omega_2^{-1} Q_2^{-1} =0.003$,
$J_1=10^{-4}$, and $J_2=1$. Comparison with Fig.~\ref{fig2} makes it
clear that the overall outline of the curves is the same as for the
self-sustained Duffing oscillator considered in Section \ref{FS}.
However, a crucial feature shows up for synchronization frequencies
around the higher-harmonic frequency $\omega_2$. In the close-ups of
Fig.~\ref{fig4}c, we see that the resonance curve develops a gap
across the direction of the frequency axis, so that the peak is cut
off from the rest of the curve in a kind of elongated ``island.''
For frequencies within the gap, the only possible solution lies on
the low-amplitude branch, where the curve acquires in turn an
up-down peak whose center and width coincide with those of the gap.

In terms of the variation of the control parameter $\phi_0$, in
Fig.~\ref{fig4}a, we see that only one solution exists for small
phase shifts, with a rapidly increasing frequency. As $\phi_0$ grows
further and the frequency approaches that of the higher-harmonic
mode, however, the growth of the frequency flattens abruptly and, at
the same time, two new solutions appear. Of these two new solutions,
the one with the lower frequency remains close to the preexisting
solution, with which it determines the frequency gap referred to in
the preceding paragraph. Eventually, well beyond the central part of
the curves, the two additional solutions disappear and the frequency
of the preexisting solution decreases. Note the sizable interval of
phase shifts ($0.6 \lesssim \phi_0 \lesssim 2.6$ for the parameters
of Fig.~\ref{fig4}) along which the preexisting solution maintains
its frequency at a practically constant level. For all these values
of our control parameter, the main-mode frequency is thus strongly
stabilized by the internal resonance. A small additional interval
with three solutions, where the frequency is again stabilized, is
found for larger values of the phase shift ($2.7 \lesssim \phi_0
\lesssim 3$) in correspondence with the up-down peak of the
resonance curve.

From the third and fourth of Eqs.~(\ref{algebres}) it can be
immediately seen that the amplitude of the higher-harmonic mode is
\begin{equation} \label{A2}
A_2 = \frac{\omega_2^2 J_2 A_1}{\sqrt{(\omega_2^2 -\Omega^2)^2+
\omega_2^2 Q_2^{-2}\Omega^2 }} .
\end{equation}
As expected for a linear oscillator, this amplitude is proportional
to the  forcing amplitude $J_2 A_1$, but is also modulated by a
frequency-dependent Lorentzian factor. This modulation implies that
$A_2$ attains significant values only when the synchronization
frequency $\Omega$ reaches the vicinity of the higher-harmonic
frequency $\omega_2$, i.e. around the frequency gap. Equation
(\ref{A2}) also shows that the width of this zone is $\Delta \Omega
\sim \omega_2 Q_2^{-1}$, and is therefore inversely proportional to
the quality factor of the higher-harmonic mode. It is within this
zone that the coupling between the two modes is most effective, and
resonant energy transfer takes place from the self-sustained main
mode.

Taking into account the results presented in Section \ref{FS},  the
stability of the periodic solutions given by Eqs.~(\ref{algebres})
can be inferred from a standard analysis of the plot in
Fig.~\ref{fig4}a, which we interpret as a bifurcation diagram for
the frequency as a function of the phase shift $\phi_0$. Assuming
that the solution branches present in both Figs.~\ref{fig2} and
\ref{fig4} have the same stability properties, we expect that the
only solution found in Fig.~\ref{fig4}a for small phase shift is
stable. The appearance of two new solutions as the phase shift grows
should be associated with a saddle-node bifurcation, creating a pair
of solutions with opposite stability. The upper branch, also present
in  Fig.~\ref{fig2}a, should correspond to the stable one. Upon
further increasing of $\phi_0$, an inverse saddle-node bifurcation
annihilating the same pair takes place  and, a little farther, a new
saddle-node bifurcation gives rise to the two solutions in the small
peak. The unstable solution of this new pair annihilates in turn
with the preexisting stable solution, while its stable partner
subsists until the phase shift attains its largest values,
constituting the rightmost branch of the bifurcation diagram.

We have verified these conclusions on the stability properties of
the periodic solutions by numerically solving the equations of
motion, with the same scheme as in Section \ref{FS}. Dots in
Fig.~\ref{fig4} show long-time numerical measurements of amplitude
and frequency with the same parameter choice as for the curves. The
control parameter $\phi_0$ varies from $\phi_0\approx 0$ to $\pi$ at
intervals of $0.01 \, \pi$. In the zones where three solutions
exist, we have integrated the equations of motion at least twice,
starting from different initial conditions, compatible with large
and small oscillation amplitudes. In the small interval
corresponding to the up-down peak, we have refined the phase-shift
sampling  to get better evidence on the stability of the three
solutions. For the sake of clarity, numerical results in this
interval are not included in Fig.~\ref{fig4}.

\begin{figure}
\includegraphics[width=0.75\textwidth]{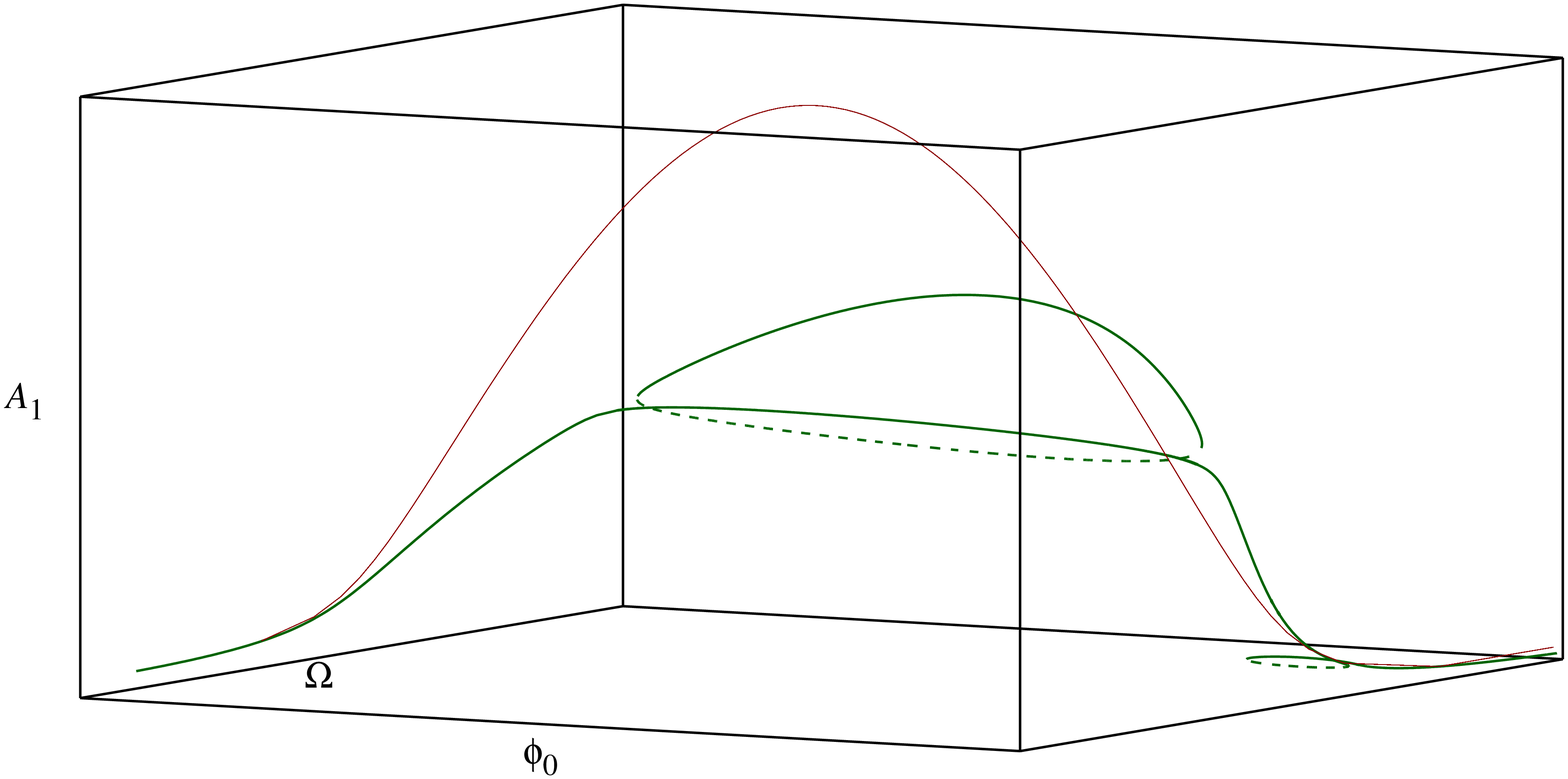} \\
Online Resource 1  (ESM{\_}1.gif): Animation with rotating
three-dimensional visualization of the internal-resonance curve in
the $(\phi_0,\Omega,A_1)$-space, as given by the solution to
Eqs.~(\ref{algebres}) with the parameters of Fig.~\ref{fig4}. The
resonance curve for a forced linear oscillator with the same $Q$ and
$f_0$ is plotted for comparison.
\end{figure}

The bifurcation scenario is clearly depicted by the plot of
frequency vs.~phase shift in Fig.~\ref{fig4}a (see inset), but may
turn out to be less obvious in the plot of amplitude vs.~phase shift
in Fig.~\ref{fig4}b, or in the resonance curve. In order to ease the
visualization of the whole picture, the Online Resource 1 contains
an animation with views in the three-dimensional space spanned by
the coordinates $(\phi_0, \Omega, A_1)$, which illustrates how the
three main plots of Fig.~\ref{fig4} connect with each other.

\subsection{Preliminary experimental results and the backbone
approximation}

As mentioned above, previous experiments on frequency stabilization
by internal resonance in clamped-clamped microoscillators were
carried out for fixed phase shift ($\phi_0 \approx \pi/2$) and
varying the amplitude of the self-sustaining force \cite{mm2}. Under
these conditions, due to the effect of the cubic force, the
oscillation frequency increases with the amplitude until it reaches
the resonance region. The results of Section \ref{IRth}, in turn,
show that the resonance can also be induced by keeping the
self-sustaining amplitude fixed and varying the phase shift.

\begin{figure} \includegraphics[width=0.6\textwidth]{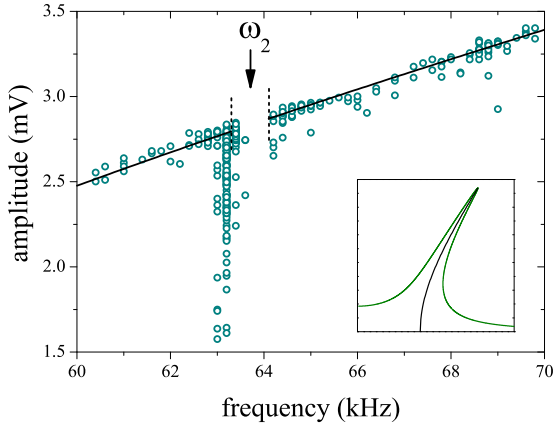}
\caption{Experimental measurements (dots) of the  amplitude $A_1$
and frequency $\Omega$ of the main oscillation mode under
phase-shift control of a self-sustained clamped-clamped
microoscillator. Experimental errors (not displayed) were estimated
in $0.015$ mV for the amplitude and $0.1$ kHz for the frequency.
Vertical dashed segments indicate the frequency gap caused by the
internal resonance, around $\omega_2 \approx 63.7 $ kHz. The solid
line is a fit of the experimental data with the backbone
approximation to our analytical model. The inset illustrates the
backbone approximation for a resonance curve with the parameters of
Fig.~\ref{fig2} ($Q=10$). The approximation improves sharply as $Q$
grows.} \label{fig5}
\end{figure}

We have performed preliminary measurements under phase-shift control
on the same kind of self-sustained microoscillators as used in
previous experiments. The phase shift is tuned by means of a
variable resistor in an all-pass filter intercalated in the
self-sustaining feedback circuit. Details of the electronics can be
found elsewhere \cite{tesisSA}. Dots in the main plot of
Fig.~\ref{fig5} stand for our measurements in the
amplitude-frequency plane, i.e.~over the resonance curve. The
amplitude is measured from an oscillating electric signal produced
by the vibrating silica bar through a capacitive transducer, and is
therefore given in millivolts. The uncertainty in the determination
of amplitude and frequency was around $0.015$ mV and $0.1$ kHz,
respectively.

As the phase shift is increased from small values, first, the
frequency and the amplitude grow as well. Then, when the frequency
reaches some $63.3$ kHz, the oscillations suddenly become irregular,
with their amplitude varying erratically over a rather wide
interval. The frequency, on the other hand, remains stable at a
rather well-defined value, insensible to the amplitude fluctuations.
Frequency stabilization is apparent at this point. Further increase
of the phase shift leads again to more stable oscillations,
regaining the regime where both the amplitude and the frequency grow
with $\phi_0$. A gap, however, has been left vacant for frequencies
between $63.3$ and $64.1$ kHz, corresponding to amplitudes between
$2.8$ and $2.9$ mV, approximately.

Our experimental results can be satisfactorily fitted by the model
presented in Section \ref{IRth}, with a suitable choice of its
parameters. To this end, it is useful to note first that the quality
factor of a clamped-clamped silica microoscillator as used in the
experiments is typically in the order of $10^4$ \cite{mm2}. This
large quality factor implies that the Duffing resonance curve is
very narrow, with its two leaning branches very close to each other.
Under these conditions, the two branches are well represented by a
single curve --sometimes called {\em backbone}  curve
\cite{Nayfeh}-- given by an approximate expression for the
interdependence between amplitude and frequency. The inset of
Fig.~\ref{fig5} shows a resonance curve with $Q=10$ and its backbone
approximation. As $Q$ grows and the curve becomes narrower, the
approximation is increasingly good.

The backbone approximation for the Duffing resonance curve is
obtained from Eqs.~(\ref{algeb}) by neglecting the contribution of
damping and of the external force. Thus, it corresponds to assuming
not only a large quality factor but also a large oscillation
amplitude. With the notation of Eqs.~(\ref{algeb}), the backbone
curve is given by $A=\sqrt{(\Omega^2-1)/\tilde \beta}$. The value of
the phase shift $\phi_0$ along the curve can then be determined from
the second of those equations: $\sin \phi_0= \Omega A/Q f_0$. In
order to apply this approximation to fit the experimental data shown
in Fig.~\ref{fig5}, we first disregard the effect of the internal
resonance, and  write
\begin{equation}
A_1= a_1 \sqrt{(\Omega /\omega_1)^2-1},
\end{equation}
where $a_1$ and $\omega_1$ are tunable parameters, respectively
given by the units of measure of $A_1$ and $\Omega$. Our estimate
yields $a_1 = (6.4 \pm 0.1) \times 10^{-2}$ mV, and $\omega_1 =
(46\pm 1)$ kHz. The line in the main plot of the figure stands for
this estimate. Assuming for our microoscillator a quality factor of
the order of $10^4$, the width of the resonance curve would fall
well inside the dispersion of the experimental data, which justifies
using the backbone approximation.

To fit the gap in the backbone curve, indicated by vertical dashed
lines in Fig.~\ref{fig5},  we now turn the attention to
Eqs.~(\ref{algebres}). Solving the two last equations for the
product $A_2 \sin \psi_2$, and replacing the result into the second
equation, makes it possible to write
\begin{equation} \label{gap}
\sin \phi_0 = \left[ 1 + \frac{\nu^4}{(\omega_2^2 -\Omega^2)^2+
\omega_2^2 Q_2^{-2}\Omega^2 }  \right] \sin \Phi_0.
\end{equation}
The constant $\nu^4$ --where $\nu$ has units of frequency-- is
proportional to the product of the coupling constants, $J_1 J_2$,
and to the ratio of the quality factors of the two involved
oscillation modes, $Q_1/Q_2$. These quantities cannot be discerned
from each other in our experiment, and $\nu$ must therefore be
considered as a single fitting parameter. The angle $\Phi_0$, in
turn, can be associated to the phase shift in the zone of the gap
when the internal resonance is absent, i.e.~for $\nu^4 \propto
J_1J_2 =0$. From our measurements of amplitude and frequency as
functions of the phase shift  (not presented here) we estimate $\sin
\Phi_0 = 0.16 \pm 0.01$.

Equation (\ref{gap}) makes it clear that the gap in the backbone
curve appears when the second term inside the square bracket in the
right-hand side is large enough as to make $\sin \phi_0 > 1$, for
which no real value of the phase shift $\phi_0$ satisfies the
equation. The Lorentzian profile of this term as a function of  the
frequency  $\Omega$ indicates that the  gap occurs around
$\omega_2$. As advanced in Section \ref{IRth}, its width $\Delta
\Omega$ is proportional to $\omega_2 Q_2^{-1}$. Locating the center
of the gap, we estimate $\omega_2 = (63.7 \pm 0.2)$ kHz. Tuning the
constant $\nu$, on the other hand, requires knowing the quality
factor $Q_2$ of the higher-harmonic mode, to which we do not have
access in the experiment. However, if --in the spirit of the
backbone approximation-- we assume that $Q_2$ is large, the
contribution of the term $\omega_2^2 Q_2^{-2}\Omega^2$ in the
denominator of the Lorentzian function can be neglected by
comparison to $(\omega_2^2 -\Omega^2)^2$. Within this assumption, we
find $\nu^4 = (\omega_2 \Delta \Omega )^2 (1-\sin \Phi_0)/\sin
\Phi_0$. Taking $\Delta \Omega = 0.8$ kHz, our estimate yields $\nu
= (10.8 \pm 0.2)$ kHz.

\section{Conclusion} \label{C}

Micromechanical devices have opened the possibility of renewed
mutual contribution between nonlinear physics and technological
applications \cite{mm1,mm4}. On one side, this minute machines
--still belonging to the realm of Newtonian mech\-anics-- often
function  within regimes where nonlinear effects play a substantial
role in the dynamics. This is particularly true for micromechanical
oscillators,  which are foreseen to be used as pacemakers in
time-keeping electronic devices, and whose large vibration
amplitudes bring them well beyond the linear elastic regime. On the
other side, well-developed techniques of MEMS fabrication can be
used to build up microscale lab equipment, where complex physical
phenomena such as the collective dynamics of large populations of
nonlinear coupled oscillators may be realized and tested.  This path
is just beginning to be explored \cite{s1,s2}.

In this paper, we have studied two technologically relevant aspects
of the self-sustained Duffing oscillator, which models a
clamped-clamped elastic bars inserted in a feedback circuit.
Firstly, following the traditional multiple-scale approach used to
study the stability of the standard Duffing equation \cite{Nayfeh},
we have shown that the self-sustained oscillator is stable along the
entire resonance curve. This result contrasts with the well-known
standard behavior, where an unstable branch develops in the middle
part of the curve. The difference resides in the fact that, in the
self-sustained oscillator, the system is controlled by fixing the
phase shift between oscillation and forcing whereas, in the standard
case, the control parameter is the forcing frequency. The situation
is reminiscent --although not fully equivalent-- to the basic
process of feedback stabilization prescribed by control theory. We
have shown that, within the multiple-scale approximation, the
equations of motion can be exactly solved, and provided numerical
evidence that validate the approximation.

In the second place, we have elaborated on a model for the internal
resonance between the main oscillation mode and a higher-harmonic
mode in a clamped-clamped oscillator. The interest of this
phenomenon lies in that it has been associated with the frequency
stabilization observed in microoscillators upon variation of the
forcing amplitude \cite{mm2}. Frequency stability under amplitude
fluctuations is an unavoidable condition for the use of
microoscillators as frequency references in time-keeping devices.
The model describes the internal resonance as a synchronization
process between the main nonlinear oscillation, represented by the
self-sustained Duffing equation, and a linear oscillator of
different frequency coupled to the main mode. Due to this
phenomenon, the Duffing resonance curve develops a gap, which is
associated to a wide phase-shift interval where the oscillation
frequency remains practically constant. As similar ``island'' effect
has recently been reported for an externally forced Duffing
oscillator coupled through displacement and velocity to a linear
oscillator \cite{Habib}. It is interesting to point out that, in the
vicinity of the gap, the amplitude-frequency interdependence is
similar to that of the standard forced Duffing oscillator at its
subharmonic resonances \cite{Nayfeh}. In order to establish the
stability properties of the model, we have performed numerical
simulations. These support a picture where the resonance occurs
through two pairs of direct-inverse saddle-node bifurcations.
Finally, we have presented preliminary experimental results which
illustrate the internal resonance in the situation where, contrary
to previous experiments, the phase shift between forcing and
oscillation was controlled. Model and experiment were fitted to each
other within a high-quality-factor approximation.

At the microscale, an important dynamical ingredient --that has
however been purposely left aside in our description-- is added by
noise. Random fluctuations of both thermal and electronic origin
become increasingly influential on the mechanics of micromachines as
these decrease in size. In the case of oscillators, noise makes it
necessary to work at larger amplitudes, which in turn increases the
effects of nonlinearities. To our knowledge, a theoretical approach
to the interplay between noise and nonlinearities focused on the
dynamics of micromechanical systems is still lacking. This seems to
be an appealing next step in the lines of the present research,
where the physics of stochastic processes should have their say.

\begin{acknowledgements}
We acknowledge financial support from ANPCyT (PICT 2011-0545),
Argentina, and enlightening discussions with Hern\'an Pastoriza,
Dar\'{\i}o Antonio, and Daniel L\'opez. Experiments were conducted
at the MEMS Laboratory of Centro At\'omico Bariloche.
\end{acknowledgements}

\end{document}